\documentclass[12pt]{article}

\usepackage{graphicx}
\graphicspath{{figures/}}
\DeclareGraphicsExtensions{.eps}

\begin{document}
\def\KAOS{{\sc Kaos}\@}
\def\PANDA{$\overline{\mbox{P}}${ANDA}\@}
\title{Hypernuclear physics\\ as seen by an experimenter}
\author{Patrick Achenbach\\
  \small Institut f\"ur Kernphysik, Joh. Gutenberg-Universit\"at,
  Mainz, Germany.}
\date{2008}
\maketitle
In the new millennium hypernuclear physics is undergoing a renewed interest,
both theoretically and experimentally. 

Hadrons and nuclei are understood as many-body systems made of quarks and gluons, bound by the strong force. Information on baryon-baryon interactions is mainly obtained from
nuclear experiments with projectiles and targets out of nucleons, 
addressing interactions in flavour SU(2) only.
The difficulties to study hyperon-nucleon ($YN$) and hyperon-hyperon ($YY$) interactions by reaction experiments are related to the practical problems in the preparation of
low energy hyperon beams and the impossibility of hyperon targets due
to the short ($\sim$ 200\,ps) life-times of hyperons. 
The investigation of a hypernucleus, where one or more nucleons have been replaced by one or more hyperons, allows addressing a rich spectrum of physics topics ranging from genuine nuclear physics to particle physics. Although fifty years have already passed since the discovery of the first hypernuclear events, studies of hypernuclei are still at the forefront of nuclear physics.  The presence of the hyperon can induce several effects on the host nucleus, like changes of both size and shape, modification of cluster structure, manifestation on new symmetries or changes of nucleon collective motions. One of the most spectacular effects, observed so far in what is called impurity nuclear physics, is the shrinking of the nucleus core. Such a behaviour can be considered a precursor of matter condensation induced by strange particles.

Only recently, it has already been
demonstrated that hypernuclei can be used as a micro-laboratory to
study $YN$ and $YY$ interactions.
In the case of $\Lambda N$ interaction, the spin-orbit term has been
found to be smaller than that for the nucleon. In a recent experiment
at BNL the spacing of the $(5/2^+,3/2^+)$ doublet in
$^{9}_{\Lambda}$Be was measured to be $(43 \pm
5)\,$keV~\cite{Tamura2005}. Although these small spin splittings can
only be observed using gamma spectroscopy, reaction spectra are
equally important because they provide the complete spectrum of
excitations. In addition, experimental data on
medium to heavy single $\Lambda$ hypernuclei have shown a much larger
spin-orbit splitting than observed in light
hypernuclei~\cite{Nagae2000}.

Hypernuclei physics, born and developed mainly in Europe, has seen a renaissance at the turn of the century. Until now, experimental information has mainly come from meson-induced reactions and most recently from coincident $\gamma$-ray spectroscopy of hypernuclei. Even though a number of new experimental techniques have been developed for the hypernuclear spectroscopy in the last decade, our knowledge is still limited to a small number of hypernuclei. The large variety of novel experimental approaches to hypernuclei will provide a wide basis for a comprehensive understanding of strange hadrons in cold nuclear systems.  
The
spectroscopy of single $\Lambda$- and double $\Lambda\Lambda$-hypernuclei will
remain one of the most valuable tools for the experimental investigation of
strangeness nuclear physics in the near future. 

\section{The hypernuclear programme at MAMI}
At the Institut f\"ur Kernphysik in Mainz, Germany, the microtron MAMI
has been upgraded to 1.5\,GeV electron beam energy and can now be used
to study strange hadronic systems~\cite{Kaiser2008}. 

Electron beams have excellent spatial and energy definitions, and
targets can be physically small and thin ($10-50$\,mg$/$cm$^2$)
allowing studies of almost any isotope.  The cross-section for
the reaction, $\sigma\sim 140$\,nb$/$sr on a $^{12}${C}
target as first measured at Jefferson Laboratory in Experiment E89-009~\cite{Miyoshi2003}, 
is small compared to strangeness exchange
n$(K^{-},\pi^{-})\Lambda$ or associated production
n$(\pi^{+},K^{+})\Lambda$. This smallness can be well compensated in electroproduction 
by the available large electron beam intensities, but often the resulting electromagnetic
background is limiting the reaction rates.

\begin{figure}
  \includegraphics[width=0.49\textwidth]{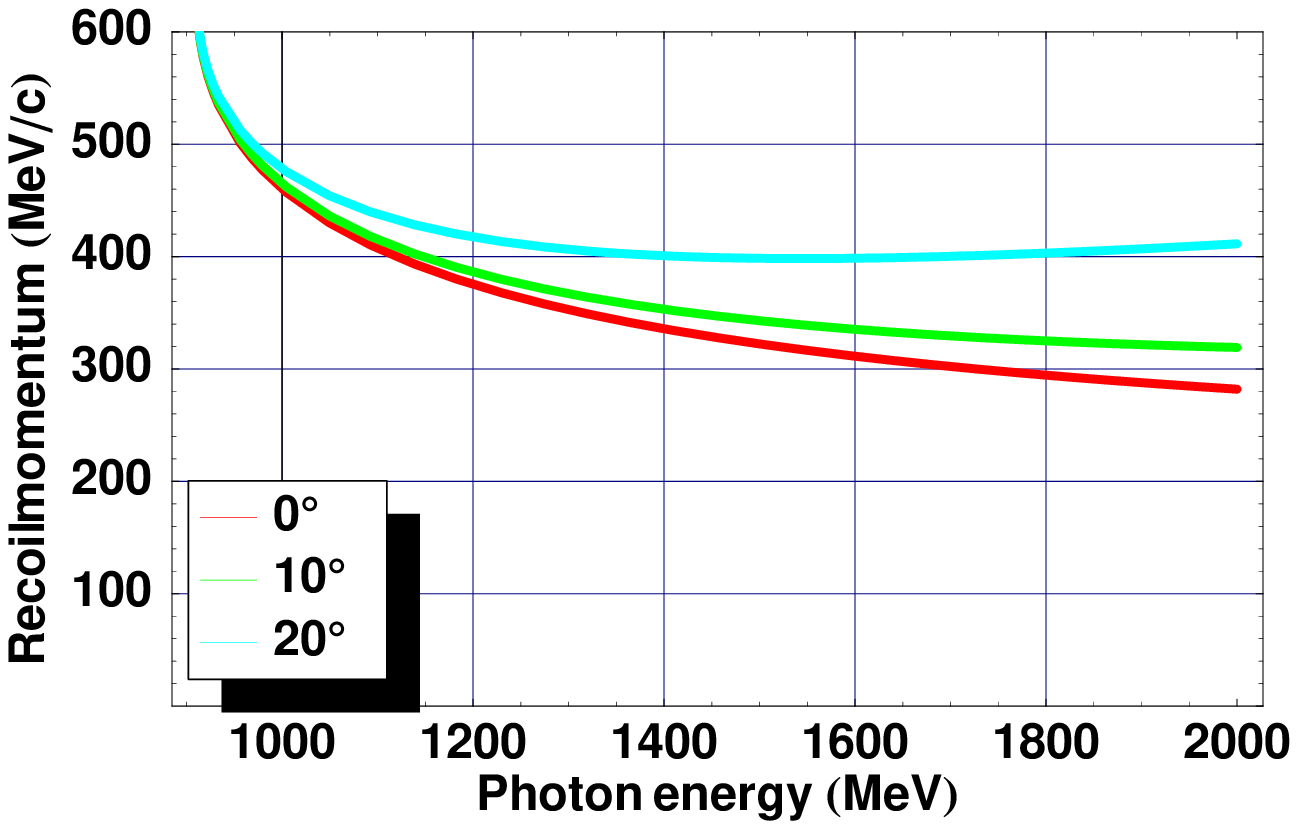}
  \includegraphics[width=0.49\textwidth]{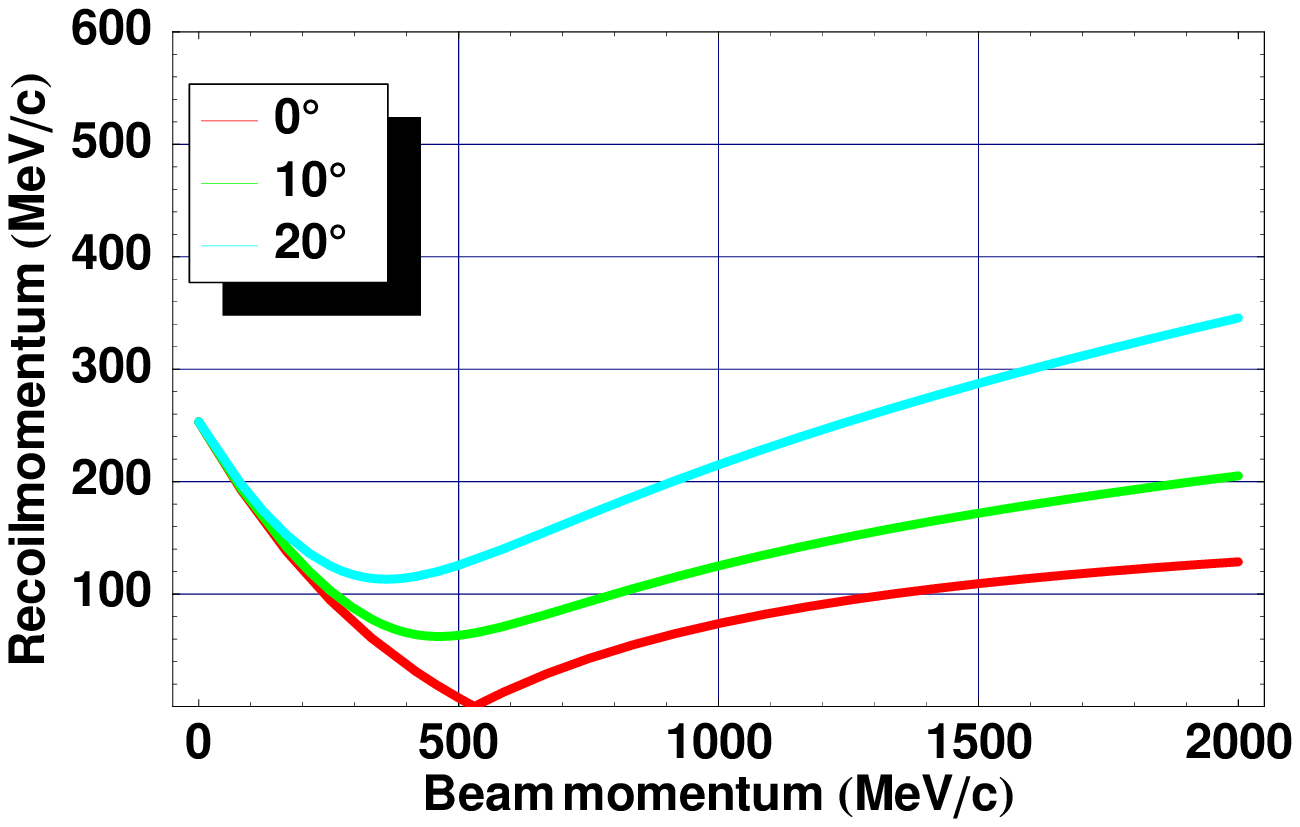}
  \caption{Recoil momentum for strangeness electro-production (left) and
  	strangeness exchange (right) reactions at three different kaon
  	angles are shown as a function of the energy of the
  	virtual photon, respectively the beam momentum. 
  	Reaction cross-sections and transition amplitudes to individual states 
  	depend strongly on the recoil momentum.}
  \label{fig:recoil}
\end{figure}

In order to produce a hypernucleus, the hyperon emerging from the
reaction has to be bound in the nucleus. Reaction cross-sections 
and transition amplitudes to individual states 
depend strongly on the transferred momentum to the hyperon.  If the
momentum transfer is large compared with typical nuclear Fermi
momenta, the hyperon will preferentially leave the nucleus.
The $(K^{-},\pi^{-})$ reaction is characterised by the existence of a
"magic momentum" where the recoil momentum of the hyperon becomes
zero as is shown Fig.~\ref{fig:recoil}. It populates, consequently,
substitutional states in which a nucleon is converted to a $\Lambda$
in the same state. The $(e,e',K^{+})$ reaction, on the other hand,
produces neutron-richer $\Lambda$ hypernuclei converting a proton to a
$\Lambda$ hyperon and transfers a large recoil momentum to a
hypernucleus. This reaction is preferable when high-spin hypernuclear
states are studied. In addition, this reaction has the unique
characteristic of providing large amplitudes for the population of
spin-flip hypernuclear states with unnatural
parities~\cite{Motoba1994}, such as $(_\nu p_{3/2}^{-1}, _\Lambda
s_{1/2})2^-$, where the spin quantum number, $J^P= 2^-$, of the
nucleon-hole $\Lambda$-particle state has maximum $J= _\nu\!l +
_\Lambda\!l + 1= 1 + 0 + 1= 2$.

\KAOS\ is a very 
compact magnetic spectrometer suitable especially for the detection of 
kaons, that was used before at GSI in a single-arm
configuration~\cite{Senger1993}. During the last years it 
was installed at the Mainz microtron MAMI in the existing spectrometer facility 
operated by the A1 collaboration~\cite{Blomqvist1998}. 
In the very near future the spectrometer
will be set-up for the first time with tracking detectors arranged in two arms, to either
side of the main dipole. 
The special kinematics for electroproduction of hypernuclei
requires the detection of both, the associated kaon and the
scattered electron, at forward laboratory angles. 
The \KAOS\ spectrometer will cover simultaneously electron scattering angles 
close to 0$^{\circ}$ and kaon scattering angles around 5$^{\circ}$ up to
15$^{\circ}$ in order to extract dynamical information from the
$K^+$ angular distribution~\cite{Achenbach-HYP06}. 

\begin{figure}
  \centering
  \includegraphics[height=0.6\textwidth,angle=90]{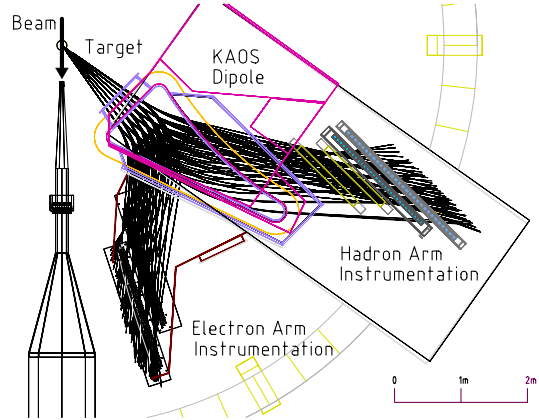}
  \caption{Overview of the \KAOS\ spectrometer of the A1 
  	collaboration at the Mainz microtron MAMI: electrons and 
  	hadrons are detected simultaneously under small scattering angles.
  	Charged 
  	particle trajectories through the spectrometer are shown by full lines.
  	The electron arm tracking detector will be located close to the electron beam.
  	High radiation levels are expected at that position.}
  \label{fig:KAOS}
\end{figure}

The \KAOS\ spectrometer's electron arm detectors will operate close to
zero degrees scattering angle and in close proximity to the electron
beam. Fig.~\ref{fig:KAOS} shows a schematic drawing of the set-up in the
spectrometer hall.  
The magnet bends the central
trajectory on both sides by $\sim$45 degrees with a momentum dispersion of 2.2\,cm$/$\%. 
The first-order focusing is realized as seen in Fig.~\ref{fig:KAOS}.
In addition to a broad
neutron spectrum high electromagnetic background levels are expected
at the detector locations. It is consequently imperative to operate 
radiation hard and intrinsically fast detectors.

While the instrumentation of the hadron arm is operational, a new
coordinate detector of the spectrometer's electron 
arm is under 
development~\cite{Achenbach-SNIC06,Achenbach2008}. 
It will consist of two vertical planes of fibre arrays ($x$ and $\theta$), 
covering an active area of $L \times H \sim$ 2000 $\times$ 300\,mm$^2$, 
supplemented by one or two horizontal planes ($y$ and $\phi$). 
The 18,432 fibres of the vertical tracking
detectors will be connected to 4,608 electronics channels with logic signals fed 
into the level-1 trigger. The track information will be used to reconstruct the target coordinates
and particle momentum, and the time information used to determine the time-of-flight of the
particle from target to the detection planes.
New front-end electronics has been
developed for the fast signals of more than 4,000 MaPMT channels of the fibre detector
in the \KAOS\ spectrometer's electron arm.

\begin{figure}
	\centering
  \includegraphics[width=0.6\textwidth]{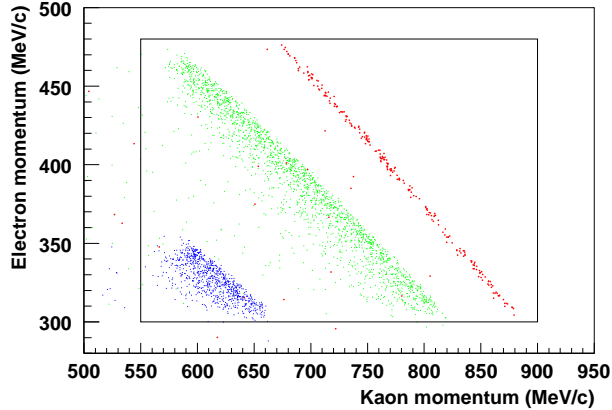}
  \caption{(color online) Simulated correlation between electron and kaon momenta,
    where $\Sigma$ (blue, left) and $\Lambda$ (green, centre) hyperons have been generated for the elementary
    production off the proton and the $^{12}_{\Lambda}$B hypernuclei (red, right) have been
    generated for a carbon target. The rectangular box indicates the simultaneous momentum acceptance
    of the \KAOS\ spectrometer in its two-arm configuration.}
  \label{fig:correlation}
\end{figure}
In Fig.~\ref{fig:correlation} the simulated correlation in electroproduction
between the electron momentum 
and the kaon momentum is plotted, where $\Lambda$ and $\Sigma$ hyperons
have been generated for the elementary production off the proton
and the $^{12}_{\Lambda}$B hypernuclei have been generated for a
carbon target. The events have been generated randomly in
phase-space and weighted by a factor for the virtual photon flux
and the modelled transition form factor. In the Monte Carlo, the
production probability was assumed to drop exponentially with
the relative momentum between $\Lambda$ hyperon and core nucleus and
typical values of $\sigma_p=$ 100\,MeV$/c$ and $k_F=$
200\,MeV$/c$ were assumed. The rectangular box in Fig.~\ref{fig:correlation}
indicates the simultaneous momentum acceptance of the \KAOS\ spectrometer. Its
large momentum acceptance covers the quasi-free process as well as the hypernuclear
production reaction. In practice, this fact will simplify the identification of the
hypernuclear events in the data sample.

It is currently planned to perform a first
experiment with two complete vertical planes of the fibre detector in the \KAOS\ spectrometer's
electron arm in 2009. The hypernuclear programme will follow as soon as the two-arm configuration
of the spectrometer is operational and the magnet optics is determined in such a way that sub-MeV 
mass resolution is possible. The latter situation is assumed to be reached in late 2009 or early 2010.

\section{The HypHI experiment}
Until recently hypernuclear spectroscopy has been restricted to the investigation of hypernuclei
close to the valley of beta-decay stability as in most experiments targets made of stable
nuclei are used with meson and electron beams. 
The recently proposed HypHI project 
(Hypernuclear spectroscopy with stable heavy ion beams and rare-isotope beams) 
is dedicated to hypernuclear spectroscopy with
stable heavy ion beams and rare isotope beams at GSI, Germany, and
FAIR, the Facility for Antiproton and Ion Research~\cite{Saito-HYP06}.
This approach has some advantages: firstly, it is
possible to investigate a number of hypernuclei simultaneously in a single experiment
and secondly the hypernuclei are created
at extreme isospins.
The observation of the $\Lambda$-hypernucleus decay modes 
offers the unique opportunity to look at the four-baryon, strangeness-changing, weak vertex. The determination of the relative weights of the different decay channels represents a long-standing puzzle.
The HypHI project is divided into four phases.
To study the feasibility of hypernuclear spectroscopy with heavy ion beams the phase 0
experiment was proposed~\cite{EA2006}, aiming at the identification of the $\pi^-$ decay channels of 
$^3_\Lambda$H, $^4_\Lambda$H and $^5_\Lambda$He produced
by $^6$Li 2\,$A$GeV beams impinging on a $^{12}$C target of 8\,g$/$cm$^2$ mass.

Hypernuclear production via heavy ion collisions is described 
by the participant-spectator model and was
first studied theoretically by Kerman and Weiss~cite{KermanWeiss1973}.
In the collisions hyperons are produced in the participant
region with a wide rapidity distribution centred around mid-rapidity. 
Hypernuclei can be formed in coalescence of hyperon(s) in the projectile
fragments, with the velocity of hypernuclei close to the
projectile velocity with $\beta >$ 0.9. Decays of
hypernuclei can be studied in-flight, and most of their decay
vertices are a few tens of a centimetre behind the
target at which hypernuclei are produced.

The experimental set-up, which will consist of an analysing dipole
magnet as well as time-of-flight (TOF) and tracking detectors, was designed to
measure the invariant mass of
particles decaying behind the target. The TOF branch will consist of a start
detector and two position-sensitive TOF walls for positive and
negative charged particles, placed behind the dipole. In addition, the scintillators will
provide energy deposit information for the charge identification of the registered particles. Three
tracking detectors made of scintillating fibres will be positioned between
target and magnet and will be used to trigger readout system on events which contain a
decay vertex behind the target. The fibre detector will also become crucial in distinguishing the
hypernuclei $^4_\Lambda$H and $^3_\Lambda$H from the background
containing $\alpha$ and $\Lambda$ particles produced at the target.

A further advantage of this approach is that hypernuclei are produced as projectile fragments
at beam rapidity that will open a way to direct measurements of hypernuclear magnetic
moments. In meson and electron beam induced experiments,
recoil momenta of produced hypernuclei are
small. Therefore, it has been impossible so far to conduct
direct measurement on hypernuclear magnetic moments
by means of spin precession in strong magnetic fields. This is one of the goals of the final project phase.

\section{The hypernuclear programme at PANDA}
\begin{figure}
  \begin{center}
    \includegraphics[width=0.38\textwidth]{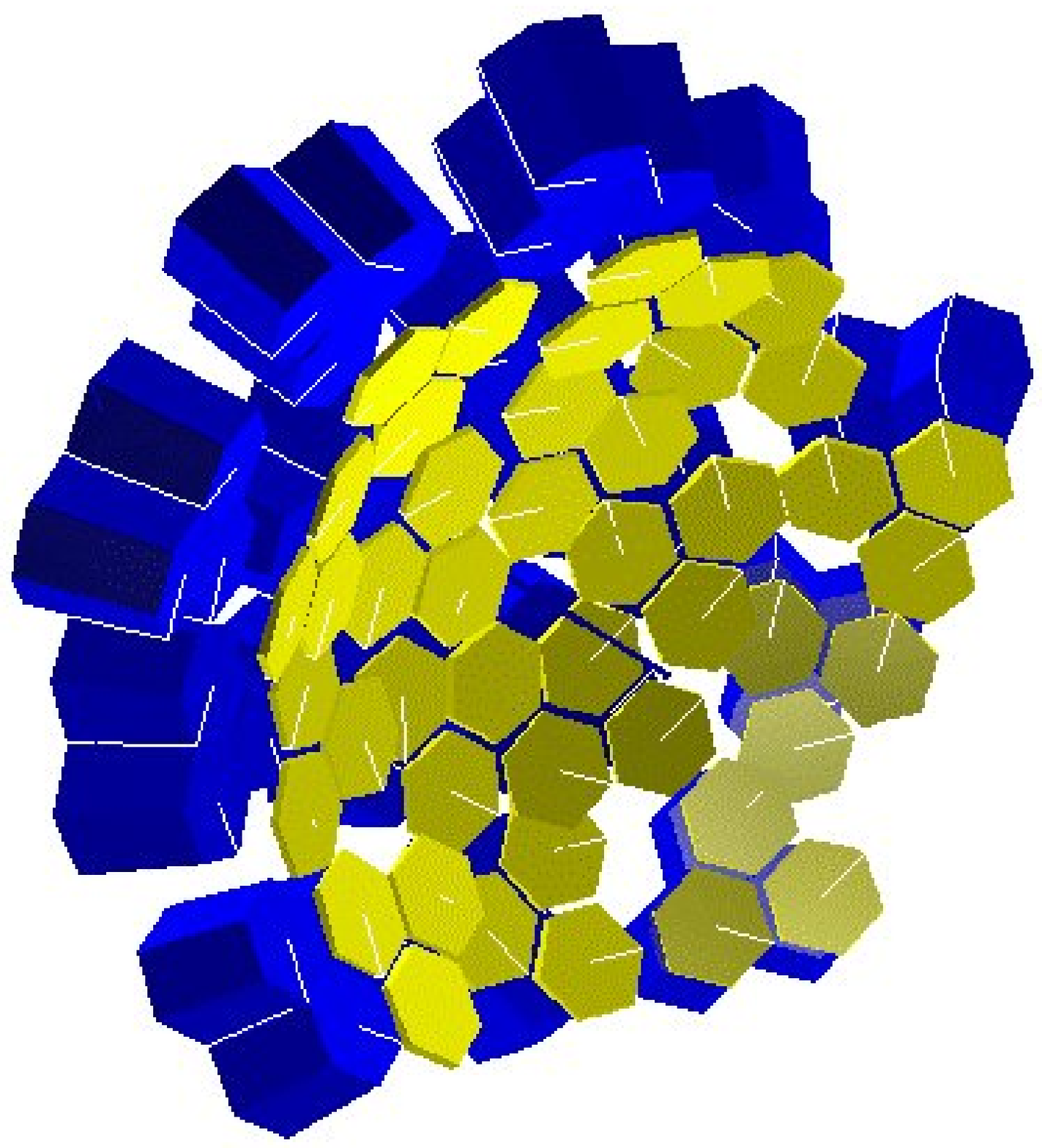}
    \includegraphics[width=0.58\textwidth]{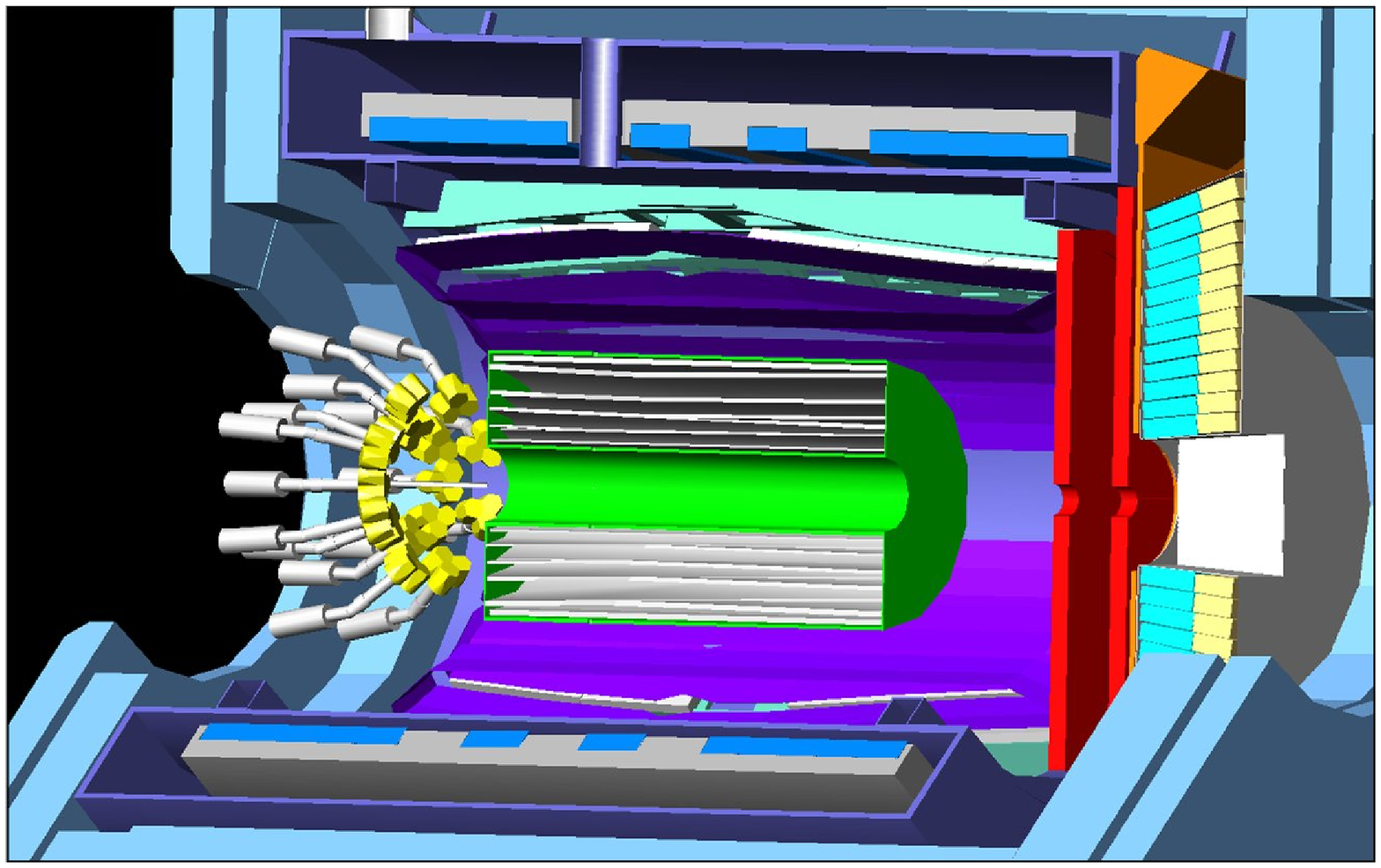}
    \caption{Simulated set-up with HPGe cluster detectors (left) situated at backward angles for
      hypernuclei experiments at \PANDA\ (right). The beam enters from
      left.}
    \label{fig:Alicia3}
  \end{center}
\end{figure}

The single hypernuclei research programme will be complemented by
experiments on multi-strange systems with \PANDA\ at the
planned FAIR facility. The \PANDA\ hypernuclear programme shall reveal the
$\Lambda\Lambda$ strong interaction strength, not feasible with
direct scattering experiments~\cite{Pochodzalla2004,PANDA2005}.
In the anti-proton storage ring HESR 
relatively low momentum $\Xi^-$ will be produced in
$\mathrm{\overline{p}p} \to \Xi^- \overline{\Xi}^+$ or
$\Xi^- \overline{\Xi}^0$
reactions. The associated $\overline{\Xi}$ will 
scatter or annihilate inside the residual
nucleus.  The
annihilation products contain at least two anti-kaons that can be used
as a tag for the reaction. Due to the large yield of hyperon-antihyperon 
pairs produced a high production rate of single and double 
hypernuclei in an active secondary
target under unique experimental conditions will be feasible.
High
resolution $\gamma$-ray spectroscopy based on high-purity germanium
(HPGe) detectors represents one of the most powerful means of
investigation in nuclear physics: the introduction of this technique
determined a significant progress in the knowledge of the nuclear
structure. Consequently, for the high resolution spectroscopy of excited hypernuclear states an
efficient, position sensitive HPGe array is foreseen. 
To
maximise the detection efficiency the detectors must be
located as close as possible to the target.
Hereby the main limitation
is the load of particles from background reactions. Most of the
produced charged particles are emitted into the forward region. Since
the $\gamma$-rays from the slowly moving hypernuclei are emitted
rather isotropically the HPGe detectors will be arranged at
backward axial angles $\theta \geq 100^\circ$.  A full simulation of
the hypernuclei detector's geometry has been completed.
Fig.~\ref{fig:Alicia3} shows the simulated $\gamma$-ray spectroscopy
set-up with several HPGe cluster detectors (each comprising 3
crystals). A small fibre barrel read-out by silicon photomultiplier has been discussed as an option
for a time-of-flight start detector to identify
hypernuclear reactions. For this 
sub-detector system the achievable time resolution at minimum detector 
mass is a main issue.

The hypernuclear physics addressed by this experiment 
is currently discussed in the upcoming ,,\PANDA\ Physics Book''.
In the planned set-up there exist many experimental challenges and several European 
research groups are working on the realisation of the detectors.
A detailed design will be available in the mid-term future. When reflecting upon the state
of the preparations for this set-up, one should be aware that the 
construction of the anti-proton storage ring and the \PANDA\ experiment
has not yet started.

\section*{Acknowledgements}
I thank the organizers for the opportunity to present and discuss practical 
issues of hypernuclear experiments with colleagues working in theoretical physics. 
I feel that such efforts are bridging the gap between the ,,cultures'' of these two fields of
research.

\bibliographystyle{elsart-num}
\bibliography{KAOS-20-09-08}

\end{document}